\documentstyle[12pt]{article}
\begin{document}
\begin{titlepage}
\vspace{60mm}
Russian Research Center "Kurchatov Institute"\\
\\
\\
\rightline{IAE-5600/2}
V.Yu. Dobretsov, V.D. Efros and Bin Shao\\
\\
\\
\\
\begin{center}
STUDY OF THE THREE-NUCLEON (e,e') LONGITUDINAL RESPONSE FUNCTION
WITH A NEW APPROACH
\vspace{120mm}
\end{center}
\centerline{Moscow-1993}
\end{titlepage}
\begin{titlepage}
PACS number(s) 25.30.Fj, 24.10.Cn, 25.10+s, 27.10+h

\vspace{30mm}

A  new  method  for  studying  the  many-body  response  functions  is
elaborated and first applied  to the $^3$He longitudinal  response. An
integral transform  of the  response function  is calculated  from the
bound-state-type  equations  for  several  versions  of the N-N force.
The  equations  are  solved  with  the  help  of  the   hyperspherical
expansion.   The  final-state  interaction  is  completely  taken into
account. The results are compared  with the integral transform of  the
experimental response function for  250 MeV/c$\leq q \leq  500$ MeV/c.
The  difference  amounts  to  20-25\%  with  the experimental response
substantially exceeding the theoretical one in the low-energy region.

\vspace{30mm}

V.Yu Dobretsov and V.D. Efros - Russian Research Center "Kurchatov Institute"\\
Bin Shao -Department of Physics, University of Pennsylvania, Philadelphia,
Pennsylvania

\vspace{30mm}

\rightline{@ Russian Research Center "Kurchatov Institute", 1993}
\end{titlepage}

Many calculations testify to the fact that the conventional form of
the  nuclear charge density is inapplicable to the description of the
elastic
form factors of  three- and four-nucleon  nuclei at $q>2.5$  fm$^{-1}$
values.   However  the  elastic  scattering  occurs  with  a quite low
probability  at  such   $q$  values  and   some  non-typical   nucleon
configurations  including  those  where  all  the  nucleons  are close
together may make  a substantial contribution.  In this connection  it
seems important to test a form of the nuclear 4-current in the
inelastic
processes and to study the ($e,e'$) response functions. This  requires
a proper account of the nuclear final-state interaction.

In this paper we present a microscopical analysis of the $^3$He
longitudinal response function $R_{l}$ with a full account of the
final-state interaction. The conventional approach  calculates the
response functions directly from the definition
\begin{equation} R(q,\omega)=\bar{\sum}_{M_0}\int df |<\psi_{f}|\hat
{O}|\psi_{0}>|^{2} \delta (E_{f}-E_{0}-\epsilon) \label{eq:defr}
\end{equation} and it requires obtaining the whole set of the
complicated final-state continuum wave functions $\psi_{f}$. (Here
$\epsilon \simeq \omega-q^2/(2AM)$ is the nuclear excitation energy.)
Therefore some model approximations always were used in such a
calculation.
We apply a new approach [1-3] which enables us to
avoid calculating $\psi_{f}$ and thereby to obtain accurate results
once the underlying nuclear dynamics is specified.

Define the reduced transition operator and the response function
\[  \tilde{O}=[\tilde{G}_{p}^{E}(Q)]^{-1}\hat{O},\,\,\,\,\,\,\,\,
\tilde{R}(q,\omega)=[\tilde{G}_{p}^{E}(Q)]^{-2}R(q,\omega). \]
Here $Q^2=q^2-\omega^2$ and [4]
$\tilde{G}_{p}^{E}(Q)=[1+Q^2/(4M^2)]^{-1/2}G_{p}^{E}(Q)$
where $G_{p}^{E}$ is the proton Sachs form factor.
We calculate the  integral transform of the response
\begin{equation}
\Phi(q,\sigma)=\int_{\epsilon_{min}}^{\infty}(\sigma+\epsilon)^{-1}
\tilde{R}(q,\omega) d\epsilon
\end{equation}
instead of the response itself. We use the conventional
single-nucleon expression for the charge density $\hat{O}$. Then to a
very good approximation one can disregard the $\omega$-dependence of
the $\tilde{O}$ operator and use the expression
\begin{equation}
\tilde{O}({\bf q})=\sum_{n=1}^{A}[\frac{1-\tau_{zn}}{2}+
\frac{G_n^E(q)}
{G_p^E(q)}\frac{1+\tau_{zn}}{2}]e^{i{\bf q}
\mbox{\boldmath $\rho$}_n}   \label {eq:op}
\end{equation}
where $\mbox{\boldmath $\rho$}_n={\bf r}_n-{\bf R}_{c.m}$. It has been
shown [1,5] that $\Phi(q,\sigma)$ can be calculated by first solving
for the {\em localized} solution to the following inhomogenous
equation
\begin{equation}
(H-E_{0}+\sigma)\tilde{\Psi}=\tilde{O}\psi_{0}  \label {eq:equ}
\end{equation}
where as in Eq. (1) $\psi_{0}$ is the ground-state wave function and
$E_{0}$
is the ground-state energy. In terms of $\tilde{\Psi}$
we have [1,5]
\begin{equation}
\Phi(q,\sigma)=<\tilde{\Psi}|\tilde{O}\psi_{0}>-\sigma^{-1}
\tilde{R}_{el}   \label {eq:phi}
\end{equation}
where $\tilde{R}_{0}$ is the elastic contribution to the response.

The solution to Eq. (\ref {eq:equ}) is much easier to obtain
than the functions $\psi_{f}$ entering Eq. (1).  Indeed, in contrast
to the latter functions there is no need to impose the complicated
large-distance boundary conditions in order to get the solution. Only
the condition that the solution vanishes at large distances is needed
in  solving Eq. (\ref {eq:equ}). Therefore methods that are used in
solving bound-state problems can be utilized here.  In particular
for many-body systems Monte-Carlo Green functions technique can be
applied.

We have two possible ways to connect our theoretical calculations
with experimental measurements. One way [5] is to
compare $\Phi(q,\sigma)$ with the same quantity
obtained from the experimental $\tilde{R}(q,\epsilon)$ using Eq. (2).
Another way [1] is to consider Eq. (2) as the integral transform and
invert to obtain theoretical $\tilde{R}(q,\omega)$ and
then compare the responses themselves.  In the
present work we use the first approach.  It is worth mentioning
that there exists a
generalization [1] of this method to exclusive reactions
including $2\rightarrow N$ reactions induced by strong interaction.

In this first calculation we use effective central $N-N$ forces [6-8]
which are supposed to act in the $s$-wave.  We supplement them with a
realistic singlet $p$-wave $N-N$ force [9]. Its contribution is
4\% at the lowest $q$ values and it is negligible at the
highest $q$ values. The contribution from the triplet $p$-wave force
is believed to be of the same size. Only the components of the
proton-proton Coulomb interaction which are diagonal in the isospin
$T=1/2, 3/2$ quantum numbers are retained in the calculation.  Even
these components change the results at most by 3\% at the lowest $q$
values considered.

Under these assumptions on the nuclear dynamics, Eq. (\ref {eq:equ})
is split
into independent sets of equations with a given orbital momentum $L$,
and isospin $T$ of the system.  It is convenient to calculate the
right-hand sides of these equations in the following way.  Since
$\psi_0$ has $L=0$ then only the components $\sim
Y_{LM_L}(\hat{\mbox{\boldmath $\rho$}}_n)$ from the expansion of
$exp(i{\bf q}\mbox{\boldmath $\rho$}_n)$ from Eq. (\ref{eq:op})
contribute to the problem for a given $L$ value.
Let ${\bf
q}$ be directed along the $z$ axis. Then only the components
with $M_L=0$ give non-zero contributions and hence only the
components of $\tilde{\Psi}$ with $M_L=0$ are different from zero. We
have
\begin{eqnarray}
(H-E_{0}+\sigma)\tilde{\Psi}_{LT}=\tilde{O}_{LT}\psi_{0},
\label {eq:eqlt}  \\
\tilde{O}_{LT}=|T><T|\sum_{n=1}^3[\frac{1-\tau_{zn}}{2}+\frac{G_n^E(q)}
{G_p^E(q)}\frac{1+\tau_{zn}}{2}]j_L(q\rho_n)Y_{L0}(\hat{\mbox
{\boldmath $\rho$}}_n),   \label {eq:oplt}\\
\Phi_{L,T}(q,\sigma)=<\tilde{\Psi}_{L,T}|\tilde{O}_{LT}\psi_0>,  \\
\Phi(q,\sigma)=4\pi\sum_{L=0}^{\infty}(2L+1)\sum_{T=1/2,3/2}\Phi_{LT}
(q,\sigma). \label {eq:suml}
\end{eqnarray}
The functions $\tilde{\Psi}_{LT}$ have the same spin $S=1/2$
as $\psi_0$.

We solve Eqs. (\ref {eq:eqlt})  by an expansion in the  hyperspherical
harmonics. Denote $K$ the hyperspherical momentum and $[f]$
the type
of symmetry of the spatial components of the basis functions.  We have
developed  a  computer  code  to   construct  complete  sets  of   the
basis functions  with arbitrary  $K,L,S,T$ and $[f]$ values using  the
Raynal-Revai   transformation   [10].   The   coefficients   of   this
transformation  are  evaluated  using  the  recurrent  formula  of the
$K\rightarrow K+2$ type  [11].  Although  the net number  of the basis
functions  with  the  same  $K,L,S,T$  and $[f]$ values grows linearly
with $K$ there  exists a possibility  indicated in Ref.  12 to specify
the basis states  in such a  way that only  two of them  contribute to
the problem in our case if one disregards the Coulomb
interaction.
It  is  because  the  nuclear  forces  that  we  use  here  only   act
in  two  N-N  orbital  states.  Only  such  states are retained in our
calculation.

We make a comment concerning calculation of $\Phi$ at small $\sigma$
values.  It is necessary to avoid large cancellations in the
right-hand side of Eq. (\ref {eq:phi}).  This is achieved if one uses
in Eq. (\ref {eq:eqlt}) for $L=0, T=1/2$  the same $K_{max}$ value as
that at calculating $\psi_0$. Then the pole terms cancel exactly.

There exists a test which enables us to check the calculation as a
whole. Namely,  the leading term  of $\Phi(q,\sigma)$ at high
$\sigma$ values behaves as $\sigma^{-1}$. This term can be calculated
independently from the sum rule,
\begin{equation}
lim_{\sigma \rightarrow \infty}\sigma \Phi(q,\sigma)=
\int_{\epsilon_{min}}^{\infty}\tilde{R}(q,\omega) d\epsilon
=\bar {\sum}_{M_0}<\psi_0|\tilde{O}^{\dag}
\tilde{O}|\psi_0>-\tilde{R}_{el}.  \label {eq:ps}
\end{equation}
This allows one to check the right-hand side of Eq. (\ref {eq:suml}).
Besides the correctness of the calculation the test allows one to
verify whether at high $\sigma$ values the results  are stable
against increasing $K_{max}$ and $L_{max}$.

The calculations are performed at $q=250$, 400 and 500 MeV/c. The
required accuracy of the calculated $\Phi$ is determined by the
accuracy of $\Phi$ extracted from experimental data. The
latter is predominantly determined by systematic errors of
the data and it is typically [13] about several per cent. We use
$L_{max}=10$ while $K_{max}=20$ for $L=0$ and $K_{max}=10$ for other
$L$ values. We have verified that these values are sufficient for our
results to converge within the experimental accuracy.

Let us first discuss the relative contributions from various $L$ and
$T$ values to the right-hand side of Eq. (\ref{eq:suml}). The
relative contributions of lower $L$ values increase as $q$ decreases
in accordance with Eq. (\ref {eq:oplt}). For example, at
$\sigma \rightarrow \infty$ the $L=0$ contribution dominates at
$q=250$ MeV/c
contributing about 60\% to the net sum while at $q=500$ MeV/c the
$L=2$
contribution dominates with the $L=0$ component contributing only
about 12 \%.
Besides, the relative $L=0$ contribution proves to increase as
$\sigma$ decreases. At $\sigma=1$ MeV and $q=500$ MeV/c the $L=0$
component contributes more than 70\%. At $\sigma=1$ MeV and $q=250$
MeV/c it contributes more than 98\%. The reason probably is as
follows.  At small $\sigma$ values the values of $\tilde{R}$ with low
excitation energies enter the integral from Eq. (2) with the highest
weights.  At low excitation energies the hyperspherical centrifugal
barrier hinders particles from being inside the reaction zone. Such a
barrier is absent when $K=0$, which value exists only for the $L=0$
component.
The effect is more pronounced at lower $q$ values since the spectrum
is shifted to lower energies.

The $T=3/2$ contribution proves to be suppressed by one or two orders
relative to the $T=1/2$ contribution at $L=0$. The reason is that
$K=0$  is forbidden for $T=3/2$. Indeed, the $K=0$ basis state is
symmetrical under particle interchange. But the symmetrical spatial
components of the final-state wave functions appear at $T=1/2$ only.
For higher $L$ values the $T=3/2$ contributions are typically
several times smaller  than the $T=1/2$ contributions as well. This
may be due to the fact that the spatially symmetrical
final-state components provide more inter-particle attraction that
increases the amplitudes of the final-state wave functions inside the
reaction zone.

We compare our results with $\Phi(q,\sigma)$ obtained from
$^{3}$He  experimental responses [13] using Eq. (2). We note that
at $q\geq 550$
MeV/c the longitudinal data of Ref. 13 become less accurate and,
besides, the sums over the data greatly exceed the theoretical
sum rule values (see Eq. (\ref {eq:ps})). We therefore only
consider the lower $q$ values. In order to perform the
integration in Eq. (2) with a sufficient accuracy and in particular
to estimate the contribution from the unavailable high-$\epsilon$
tails of the spectra we approximated the spectra by the following
analytical expressions: $a(\omega-\omega_{thresh})^{1/2}$ in the low
$\omega$ region $\omega_{thresh}\leq\omega\leq\omega_1$,
$\sum_{n=0}^{N_{b}}b_n\omega^n$ in the region of the peak
$\omega_1\leq\omega\leq\omega_2$ and
$\sum_{n=0}^{N_{c}}c_n\omega^{-(\alpha+n)}$ in the region
beyond the peak. The parameters $a,b_n,c_n,\alpha$ and $\omega_2$ are
chosen from the least-square procedure.  Additional requirements of
continuity of the fitting spectra and their first derivatives at
$\omega_1$ and $\omega_2$ points are imposed. It turns out that a
good description at a quite wide ranges
$\omega_2\leq\omega\leq\omega_{max}$ of the spectra beyond the peaks
are provided with a single term $\sim \omega^{-\alpha}$. (In case,
say, exponentially decreasing tail-terms the description is worse.)
The best $\alpha$ values range between about 4 and 5 for all the $q$
values considered. Similar $\alpha$ values were found [14] in the
$^4$He case. We extrapolate the fits obtained  beyond the
$\omega_{max}$ values in order to take into account the contributions
from the unavailable tails of the spectra.  These contributions prove
to be quite small in our case. They reach the maxima at high $\sigma$
values where they are between 1 and 2\%.

Four versions of central $N-N$ forces acting in the s-wave are
used in our calculation. These include the S2 and S3 potentials
from Ref. 6, the MT(I+III) potential from Ref. 7 and the EH potential
from Ref. 8.  The S2 potential and the MT(I+III) potential reproduce
the $N-N$ low-energy properties and $s$-wave $N-N$ phases up to high
energies. The S3 potential fits the low-energy data and yields nearly
correct values for the binding energies and rms radii of $^3$He and
$^4$He. The EH potential fits the $s$-wave $N-N$ phases up to high
energies as well but it does not reproduce properly the low-energy
$N-N$ data (see Ref. 6).

The results of our calculation  for $q$=250 MeV/c and 500 MeV/c are
shown in Fig. 1 and 2 respectively.  For $q$=400 MeV/c the results
are similar. We plot the quantity
$[\Phi_{theor}(q,\sigma)-\Phi_{exp}(q,\sigma)]/\Phi_{exp}(q,\sigma)$
against $\sigma$. At the lowest $\sigma$ values this quantity amounts
to -25\% for the S2 potential and -21\% for the MT(I+III) potential
in case of $q=250$ MeV/c and -(22-23)\% for both potentials in case
of $q=500$ MeV/c.  For the more phenomenological S3 potential it
equals -17\% at $q=250$ MeV/c and -24\% at $q=500$ MeV/c for the same
$\sigma$ values. For the EH potential the corresponding numbers are
-8\% and -20\%. (The drawback of the EH force in reproducing the
$N-N$ data is pointed out above.) At higher  $\sigma$ values the
differences between $\Phi_{exp}$ and $\Phi_{theor}$ monotonocally
decrease and they do not exceed 6.5\% in their absolute values for
all the potentals as $\sigma$ goes to infinity. (At $q=250$ MeV/c
they are about 2\% in cases of S2 and MT potentials. The differences
in 6\% in case of $q=500$ MeV/c are perhaps connected in part with
some systematic errors of the data at high $q$ values, cf. above.)
The results obtained indicate that the theoretical response functions
are lower than the experimental ones at the low-energy wings. The
deviation from experiment obtained may be caused by some deficiencies
of the adopted conventional description of the nuclear states or, in
principle, by some deficiencies of the conventional nuclear charge
density operator. Further investigations are planned in this
connection. In any case the results presented here show that a
microscopic
study     of     the     nuclear     ($e,e'$)     response   functions
with   the   final-state   interaction   fully   taken   into  account
is            accessible            and  fruitful.

We are very indebted to C. Marchand for providing us with the
experimental data. Helpful discussions with G. Do Dang and
Yu. E. Pokrovsky are gratefully acknowledged.

\vspace{20mm}

{\bf References}\\
\\
$[1]$ V. D. \'Efros, Yad. Fiz. {\bf 41}, 1498 (1985) [Sov. J. Nucl.
Phys. {\bf 41}, 949 (1985)].\\
$[2]$ V. D. Efros, W. Leidemann and G. Orlandini, Few-Body Syst.,
1993, in print.\\
$[3]$ V. D. \'Efros, Yad. Fiz. {\bf 56}, N 7 (1993) [Sov. J. Nucl.
Phys.].\\
$[4]$ J. L. Friar, Phys. Lett. {\bf 43B}, 108 (1973); D. R. Yennie,
M. Levy and D. G. Ravenhall, Rev. Mod. Phys. {\bf 29}, 144 (1957)\\
$[5]$ V. D. \'Efros, Ukr. Fiz. Zh. {\bf 25}, (1980) 907
[Ukr. Phys. J.].\\
$[6]$ I. R. Afnan and Y. C. Tang, Phys. Rev. {\bf 175}, 1337 (1968).\\
$[7]$ R. A. Malfliet and J. A. Tjon, Nucl. Phys. {\bf A127}, 161
(1969).\\
$[8]$ H. Eikemeier and H. H. Hackenbroich, Z. Phys. {\bf 195}, 412
(1966).\\
$[9]$ R. de Tourreil, B. Rouben and D. W. L. Sprung, Nucl. Phys.
{\bf A242}, 445 (1973).\\
$[10]$ J. Raynal and J. Revai, Nuovo Cim. {\bf A68}, 612 (1970).\\
$[11]$ Ya. A. Smorodinskii and V. D. \'Efros, Yad. Fiz. {\bf 17},
210 (1973) [Sov. J. Nucl. Phys. {\bf 17}, 107 (1973)]. \\
$[12]$ V. D. \'Efros, Yad. Fiz. {\bf 15}, 226 (1972) [Sov. J. Nucl.
Phys. {\bf 15}, 128 (1972)].\\
$[13]$ C. Marchand et al., Phys. Lett. {\bf 153B}, 29 (1985) and private
communication.\\
$[14]$ A. Yu. Buki et al., preprint IAE-5397/2, 1991; V. D. Efros,
Few-Body Syst.
\pagebreak
Suppl. 6, 506 (1992).\\
\centerline{CAPTION TO FIGURES}

\vspace{10mm}

FIG. 1. The relative deviation of the calculated transformant
$\Phi_{theor}(q,\sigma)$ of the longitudinal $^3$He response function
(see Eq. (2)) from that obtained from the experimental data [13].
1-S2 potential, 2-MT(I+III) potential, 3-S3 potential,
4-EH potential
(see the text).  The $q$ value is 250 MeV/c.

\vspace{10mm}

FIG. 2. Same as Fig. 1 but for $q=500$ MeV/c.
\end{document}